\documentclass[final,12pt]{clear2023_arxiv} % Anonymized submission
\usepackage{hyperref}
\newtheorem{problem}{Problem}
\usepackage{multirow}
\usepackage{multicol}
\usepackage{mathtools}

\DeclareMathOperator*{\argmin}{arg\,min} 

\usepackage[utf8]{inputenc}
\usepackage{pgfplots}
\usepgfplotslibrary{groupplots,dateplot}
\usetikzlibrary{patterns,shapes.arrows}
\pgfplotsset{compat=newest}
\usepackage[utf8]{inputenc} % allow utf-8 input
\usepackage[T1]{fontenc}    % use 8-bit T1 fonts
\usepackage{hyperref}       % hyperlinks
\usepackage{url}            % simple URL typesetting
\usepackage{booktabs}       % professional-quality tables
\usepackage{amsfonts}       % blackboard math symbols
\usepackage{nicefrac}       % compact symbols for 1/2, etc.
\usepackage{microtype}      % microtypography
\usepackage{xcolor}         % colors

\usepackage{caption}
\usepackage{subcaption}

\newtheorem{note}{Note}
\usepackage{algorithm}
\usepackage[noend]{algpseudocode}
\usepackage{tikz}
\usetikzlibrary{arrows}
\usepackage{listings}
\def\munderbar#1{\underline{\sbox\tw@{$#1$}\dp\tw@\z@\box\tw@}}
\usepackage{hyperref}
\usepackage[capitalize,noabbrev]{cleveref}
\crefname{section}{Section}{sections}
\crefname{problem}{Problem}{problems}
\usepackage{amssymb}

\def\Var{{\textrm{Var}}\,}
\def\E{{\mathbb{E}}\,}
\def\P{{\mathbb{P}}\,}

\title[Leveraging Causal Graphs for Blocking in Randomized Experiments]{Leveraging Causal Graphs for Blocking in Randomized Experiments}
\usepackage{times}
\clearauthor{%
 \Name{Abhishek Kumar Umrawal} \Email{aumrawal@purdue.edu}\\
 \addr School of Industrial Engineering, Purdue University\\
 \addr Department of Computer Science and Electrical Engineering, University of Maryland, Baltimore County
}

\begin{document}

\maketitle
\begin{abstract}
    \textit{Randomized experiments} are often performed to study the causal effects of interest. \textit{Blocking} is a technique to precisely estimate the causal effects when the experimental material is not homogeneous. It involves stratifying the available experimental material based on the covariates causing non-homogeneity and then randomizing the treatment within those strata (known as \textit{blocks}). This eliminates the unwanted effect of the covariates on the causal effects of interest. We investigate the problem of finding a \textit{stable} set of covariates to be used to form blocks, that minimizes the variance of the causal effect estimates. Using the underlying causal graph, we provide an efficient algorithm to obtain such a set for a general \textit{semi-Markovian} causal model.
\end{abstract}


\section{Introduction}

\subsection{Motivation} \label{sec:motivation}
Studying the \textit{causal} effect of some variable(s) on the other variable(s) is of common interest in social sciences, computer science, and statistics. However, a mistake that people usually make is confusing the causal effect with an \textit{associational} effect. For instance, if high levels of bad cholesterol and the presence of heart disease are observed at the same time, it does not mean that the heart disease is caused by high levels of bad cholesterol. The question is then how do we get to know if one variable causes the other? If the answer is yes, then what is the direction (positive or negative) and what is the magnitude, of the causal effect? \citet{fisher1926arrangement} provided the framework of \textit{randomized experiments} to study the causal effect, where the variable whose causal effect is to be studied (known as \textit{treatment} or cause), is randomized over the available experimental units (like humans, rats, agricultural plots, etc.) and changes in the variable on which the causal effect is to be studied  (known as \textit{response} or effect), are recorded. A statistical comparison of values of the response with or without the treatment can therefore be done to study the existence, direction, and magnitude of the cause-effect relationship of interest.

Randomized experiments work on three basic principles, viz., \textit{randomization}, \textit{replication}, and \textit{local control} \citep{fisher1926arrangement}. Randomization states that the assignment of the treatment has to be random, and replication states the treatment should be given to multiple but homogeneous units, i.e., there are multiple observations of the effect variable for both with and without the treatment. Therefore, as long as the available experimental units are homogeneous (for instance, the fertility of all the agricultural plots is the same, the responsiveness of all the humans is the same for the drug, etc.), then a `good' randomized experiment can be carried out using the first two principles, viz., randomization and replication, which gives rise to  \textit{completely randomized design} (CRD). However, when the available experimental units are not homogeneous, i.e., some attributes of experimental units (known as \textit{covariates}) differ from each other, then the causal effect may get influenced by the covariates causing non-homogeneity (like fertility, responsiveness, etc). The remedy to this problem is the third principle of randomized experiments, i.e., local control (also known as \textit{blocking}), which states to stratify the available experimental units based on that covariate(s) causing non-homogeneity, and then randomize the treatment within those strata to eliminate the effect of the covariates. These strata are called \textit{blocks}, and this third principle along with the first two gives rise to  \textit{randomized block design} (RBD). %Blocking tries to control/eliminate the variability in the response attributed through the covariates and leads to a reduction in the variance of the estimate of the causal effect of interest.
Blocking tries to control/eliminate the variability in the response attributed through the covariates and leads to a more \textit{precise} estimation of the causal effect of interest. Precision is defined as the inverse of the variance of the estimate of the causal effect of interest.

In this paper, we focus on the problem of deciding which covariates to be used for forming the blocks while performing a randomized experiment. We consider a non-parametric setting and assume that we have access to the underlying causal structure \citep{pearl2009causality}. We provide an efficient algorithm to obtain a \textit{stable} set of covariates to be used to form blocks for a general semi-Markovian causal model. The term stable refers to the idea of finding and excluding the covariates that one should never use to create blocks. The notion and the method of achieving this stability are discussed in \cref{sec:stable_blocking}.

\subsection{Literature Review}
Statistics literature \citep{yates1935complex, kempthorne1952design, kempthorne1955randomization, cochran1948experimental} presents profound discussions on relative efficiency (ratio of the variances of causal effects) of RBD over CRD. It can be seen that in general, RBD is always more or at least as efficient as CRD. Therefore, the natural question is how do we do blocking in an intelligent manner such that maximum gain in precision can be attained? More specifically, we are interested in answering the following question. Can we decide which covariates to be used for forming blocks using the causal structure (diagram) of all the variables, viz., treatment, response, and covariates (observed and unobserved) provided the causal structure is given? 

In the past, efforts have been made to address the problem of efficient adjustments in causal graphical models under static interventions \citep{henckel2019graphical, rotnitzky2020efficient} and under dynamic interventions \citep{10.1093/biomet/asab018}. In contrast to these, we provide an efficient algorithm to obtain the set of covariates to be used for forming blocks for a general semi-Markovian causal model. %However, they do not provide a general algorithm to address the problem for a general causal graph. In contrast, we provide an efficient algorithm to obtain the set of covariates to be used for forming blocks for a general semi-Markovian causal model. 
citet{cinelli2020crash} provided a good discussion on good and bad controls for different causal graphs and emphasized the importance of good adjustments in causal estimation. Some other partially related works include \citep{kallus2018optimal} which discusses the problem of experimental design in terms of variance reduction and describes the functional assumptions placed on the potential outcome by blocking, \citep{harshaw2019balancing} which shows that discrepancy minimization (difference in means between treatment and control covariate groups) implicitly performs ridge regression and \citep{li2019rerandomization} talks about re-randomization and regression adjustment.

In this paper, we focus on deciding what covariates should form the basis of forming blocks. However, a problem that follows our problem is, how to efficiently form blocks by grouping the experimental units which are close to each other based on the chosen covariates. \citet{moore2012multivariate,higgins2016improving} focus on the latter problem.

\subsection{Contribution}
In this paper, we formalize the problem of finding a \textit{stable} set of covariates to be used to form blocks, that minimizes the variance of the causal effect estimates. We leverage the underlying causal graph and provide an efficient algorithm to obtain such a set for a general \textit{semi-Markovian} causal model.

\subsection{Organization}
The rest of the paper is organized as follows. \Cref{sec:prob_formulation} provides some background and formulates the problem of interest. \Cref{sec:methodology} discusses our methodology that includes motivating examples, key observations, important lemmas, main results, and the final algorithm. \Cref{sec:experiments} provide experimental evaluations, and \Cref{sec:conclusion} concludes the paper and provides future directions.
\section{Background and Problem Formulation} \label{sec:prob_formulation}

%Consider a \textit{structural causal model} \citetp{pearl2009causality} $M=\{U,V,F\}$, where $U$ is the set of exogenous variables, $V$ is the set of endogenous variables and $F$ is the set of functions controlling the causal structure. 

In this section, we provide some background and formulate the problem of interest in this paper.

Let $\mathcal{V}$ be the set of observed random variables and $\mathcal{U}$ be the set of unobserved random variables. Let $\mathcal{G}$ be a directed acyclic graph that depicts the causal relationships among the random variables in $\mathcal{U}$ and $\mathcal{V}$. Formally, we are assuming that the probability distribution of the random variables in $\mathcal{U}$ and $\mathcal{V}$ factorizes according to $\mathcal{G}$ (the causal Markov condition \citep{geiger1990logic}). Furthermore, we assume that no variables other than the ones that are $d$-separated \citep{pearl2009causality} in $\mathcal{G}$ are independent, i.e. independence in the joint probability distribution of the random variables in $\mathcal{U}$ and $\mathcal{V}$ is encoded by the causal graph (the causal faithfulness condition \citep{meek1995strong}).

We are interested in studying the causal effect of $X\in\mathcal{V}$ (treatment) on $Y\in\mathcal{V}$ (response). Studying the causal effect of $X$ on $Y$ means learning the (interventional) probability distribution, $\mathbb{P}(Y|do(X))$ \citep{pearl2009causality}.

\subsection{Gain in Precision due to Blocking} \label{sec:gain_in_precision}
Consider a simple setting where $X$ is an indicator random variable defined as
\begin{equation}
X = 
\begin{cases} 
    1, & \text{if the treatment is applied}, \\
    0, & \text{otherwise,}
\end{cases}
\end{equation}
and $Y$ is a real-valued random variable.

Let there be $n$ experimental units available to study the causal effect of $X$ on $Y$ by performing a randomized experiment. Let $\mathcal{Z}$ denote the set of covariates that vary across the experimental units. 

We want to demonstrate the gain in precision due to blocking. Thus, we first ignore these covariates and perform a CRD as follows. We randomize $X$ over the available experimental units, i.e., randomly assign a unit to either have the treatment ($X=1$) or not ($X=0$) and then observe the response. Let $(X_i,Y_i)$ denote the treatment and response pair for the $i$th experimental unit. %As different units are identical and we have randomized the treatment, $X_i$ and $Y_i$ are nothing but i.i.d. copies of $X$ and $Y$, respectively. In other words, $\{(X_i,Y_i): i=1,\dots,n\}$ is a random sample from ($X,Y$). 
Define $I_x:=\{i:X_i=x\}$ and $n_x:=|I_x|$. 

%denote the set of indices of units where the value of treatment, and $I_1:=\{i:X_i=0\}$ denote the set of indices of units where the treatment is not applied. Further, define $n_1:=|I_1|$ and $n_0:=|I_0|$.

%For demonstrating 
To demonstrate the gain in precision due to blocking, we focus on estimating $\mathbb{E}(Y|do(X))$. For simplicity, define $Y(x) := Y|do(X=x)$. For our simple setting, we have only two possibilities of that conditional expectation, viz. $\mathbb{E}(Y(1))$ and $\mathbb{E}(Y(0))$.  Similar to \citet{heckman1992randomization}, \citet{clements1994making}, and \citet{heckman1995assessing}, we define the \textit{effect of treatment} (ET) or \textit{average treatment effect} (ATE) in the population as
\begin{align}
\beta := \E(Y(1)) - \E(Y(0)). \label{eq:beta}
\end{align}

A natural non-parametric estimator of $\E(Y(x))$ is the corresponding sample average, given as
\begin{align}
\bar{Y}(x) &:= \frac{1}{n_x}\sum_{i\in I_x}Y_{i}.
\end{align}
Therefore, we can estimate $\beta$ as
\begin{align}
\hat{\beta} &= \bar{Y}(1) - \bar{Y}(0)
\end{align}

It can be shown that $\mathbb{E}(\hat{\beta}) = \beta$, i.e., $\hat{\beta}$ is \textit{unbiased} for $\beta$. The variance of $\hat{\beta}$ is given as follows.
\begin{align}
    \Var(\hat{\beta}) = 
      \mathbb{E}_\mathcal{Z}\left(\frac{\Var(Y(1)|\mathcal{Z})}{n_{1,z}} + \frac{\Var(Y(0)|\mathcal{Z})}{n_{0,z}} \right) + \mathbb{E}_\mathcal{Z}(\beta(\mathcal{Z})-\beta)^2, \label{eq:var_under_no_blocking}
\end{align}
where $\beta(\mathcal{Z}):= \E(Y(1)|\mathcal{Z}) - \E(Y(0)|\mathcal{Z})$ is called the \textit{$\mathcal{Z}$-specific causal effect} or \textit{conditional average treatment effect} (CATE). For proofs of unbiasedness and variance of $\hat{\beta}$, refer to \cref{app:proofs_hat_beta}.

We now perform an RBD as follows. We first stratify the experimental units such that units within each stratum (known as a block) are identical, i.e., the covariates in $\mathcal{Z}$ remain the same with a block. We next randomize $X$ over the experimental units within each block and then observe the response. Let $(X_i,Y_i,\mathcal{Z}_i)$ denote the treatment, response, and covariates triplet for the $i$th experimental unit. Define $I_{x,z}:=\{i:X_i=x,\mathcal{Z}_i = z\}$ and $n_{x,z}:=|I_{x,z}|$. 

Following \citep{petersen2006estimation}, ET defined in \eqref{eq:beta} can be re-written as 
\begin{align}
    \beta &= \E_\mathcal{Z}\left(\E(Y(1) | \mathcal{Z}\right) - \E_\mathcal{Z}\left(\E(Y(0) | \mathcal{Z}\right),\\
    &= \E_\mathcal{Z}\left(\E(Y(1) | \mathcal{Z}) - \E(Y(0) | \mathcal{Z})\right), \\
    &= \sum_{\mathcal{Z}} \left(\E(Y(1) | \mathcal{Z}) - \E(Y(0) | \mathcal{Z}) \right) \P(\mathcal{Z}) \label{eq:beta_z}
\end{align}
A natural non-parametric estimator of $\E(Y(x)|\mathcal{Z})$ is the corresponding sample average, given as
\begin{align}
    \bar{Y}(x)|\mathcal{Z} &:= \frac{1}{n_{x,z}}\sum_{i\in I_{x,z}} Y_i
\end{align}
Define $n_z:=n_{1,z} + n_{0,z}$. $\P(\mathcal{Z})$ can be estimated as
\begin{align}
    \hat{\P}(\mathcal{Z}) &:= \frac{n_z}{n} 
\end{align}
Therefore, we can estimate $\beta$ as
\begin{align}
\hat{\beta}_\mathcal{Z} &= \frac{1}{n}\sum_{\mathcal{Z}} n_{z} \left(\bar{Y}(1) | \mathcal{Z} - \bar{Y}(0) | \mathcal{Z} \right) \label{eq:hat_beta_z}
\end{align}
It can be shown that $\mathbb{E}(\hat{\beta}_\mathcal{Z}) = \beta$, i.e., $\hat{\beta}_\mathcal{Z}$ is \textit{unbiased} for $\beta$. The variance of $\hat{\beta}_\mathcal{Z}$ is given as follows.
\begin{align}
    \Var(\hat{\beta}_\mathcal{Z}) = 
      \mathbb{E}_\mathcal{Z}\left(\frac{\Var(Y(1)|\mathcal{Z})}{n_{1,z}} + \frac{\Var(Y(0)|\mathcal{Z})}{n_{0,z}} \right) + \mathbb{E}_\mathcal{Z}\left(\sum_{\mathcal{Z}}\beta(\mathcal{Z})\hat{\P}(\mathcal{Z}) - \beta \right)^2, \label{eq:var_under_blocking}
\end{align}
For proofs of unbiasedness and variance of $\hat{\beta}_\mathcal{Z}$, refer to \cref{app:proofs_hat_betaz}.

As $\hat{\mathbb{P}}(\mathcal{Z})$ is as an unbiased estimate of $\mathbb{P}(\mathcal{Z})$, $\sum_{\mathcal{Z}}\beta(\mathcal{Z})\hat{\P}(\mathcal{Z}) = \mathbb{E}_\mathcal{Z}(\beta(\mathcal{Z})) = \beta$. Hence, the second term on the right in \eqref{eq:var_under_blocking} is zero. By comparing \eqref{eq:var_under_no_blocking} and \eqref{eq:var_under_blocking}, we observe that $\Var(\hat{\beta}_\mathcal{Z}) \le \Var(\hat{\beta})$, i.e., blocking improves the precision of the estimate of the population average causal effect. %, $\beta$.

\begin{note}
\textup{Blocking is guaranteed to reduce the variance of the estimates of the causal effects under the condition the experimental units are homogeneous within each block and heterogeneous across different blocks. If the blocks are formed using some covariates that do not ensure this then blocking may, in fact, increase the variance. Hence, in practice, blocking involves a trade-off as it partitions the experimental units and thus reduces the sample size for estimating the within-block causal effect which increases the variance but may reduce the inherent variability of the response within each block which can reduce the variance. %Hence, in practice forming blocks using meaningless variables may, in fact, increase the variance.
}
\end{note}

\subsection{Challenges with Blocking} \label{sec:challenges}
We observed that in experimental studies under non-homogeneity of the experimental units, blocking improves the precision of causal effect estimation. However, the practical difficulty with blocking is that depending on the number of covariates and the number of distinct values (denoted as $v(\cdot)$) of each covariate, the number of blocks can be very large. For covariates in the set $\mathcal{Z}$, we need to form $\prod_{Z \in \mathcal{Z}}v(Z)$ different blocks. For example, if we want to study the effect of a drug on curing some heart disease where the subjects under consideration have the following attributes (which can potentially affect the effectiveness of the drug). Gender: Male and Female, i.e., $v(\text{Gender})=2$, Age: $<$25, 25-45, 45-65, $>$65, i.e., $v(\text{Age})=4$, Weight: Underweight, Normal, Overweight, Obese I, Obese II, i.e., $v(\text{Weight})=5$, Blood Pressure: Normal, Prehypertension, Hypertension I, Hypertension II, Hypertensive Crisis, i.e., $v(\text{Blood Pressure})=5$, and Bad Cholesterol: Optimal, Above Optimal, Borderline High, High, Very High, i.e., $v(\text{Bad Cholesterol})=5$. Thus, we need to form $2\times4\times5\times5\times5=1000$ blocks. Performing a randomized experiment with a large number of blocks can be very costly. Sometimes, it may not be feasible as the number of blocks can be larger than the number of subjects. This would cause some of the blocks to be empty. For instance, there may not be any male subjects under the age of 25 who are Obese II with a Hypertensive Crisis and Optimal Cholesterol level. 

%For further discussion on the challenges and importance of blocking refer to \cite{higgins2016improving} and \citet{burger2020importance}, respectively. 
Other than the economic aspects and some blocks being empty, there are reasons why some variables should never be used for forming blocks. Refer to \cref{sec:stable_blocking} for details.

% \begin{enumerate}
% \item Gender: Male and Female, i.e., $v(\text{Gender})=2$,
% \item Age: $<$25, 25-45, 45-65, $>$65, i.e., $v(\text{Age})=4$,
% \item Weight: Underweight, Normal, Overweight, Obese I, Obese II, i.e., $v(\text{Weight})=5$,
% \item Blood Pressure: Normal, Prehypertension, Hypertension I, Hypertension II, Hypertensive Crisis, i.e., $v(\text{Blood Pressure})=5$, and
% \item Bad Cholesterol: Optimal, Above Optimal, Borderline High, High, Very High, i.e., $v(\text{Bad Cholesterol})=5$.
% \end{enumerate}

%Furthermore, the cost of forming blocks may differ across covariates. Therefore, some covariates may be statistically very relevant but still economically very expensive. 

\subsection{Problem Statement}
So far, we observed that blocking improves the precision of causal effect estimation. However, in situations when the number of blocks to be formed is very large, blocking becomes costly and/or infeasible. One possible way to reduce the number of blocks is to form blocks using some but not all covariates. But the question is which covariates should be preferred over others while forming blocks? One possible way is to select the (smallest) set of covariates that leads to maximum precision in causal effect estimation. In the context of estimating the effect of treatment in the population, $\beta$, (discussed in \cref{sec:gain_in_precision}), it means finding the smallest set, $\mathcal{Z}$, that minimizes $\Var(\hat{\beta}_\mathcal{Z})$.

We next formalize the problem of interest in this paper as follows. We are given a directed acyclic causal graph, $\mathcal{G}$, that depicts the causal relationships among some observed variables, $\mathcal{V}$, and unobserved variables, $\mathcal{U}$. We are interested in studying the causal effect of $X\in\mathcal{V}$ (treatment) on $Y\in\mathcal{V}$ (response) by performing a randomized block experiment. The set of observed covariates is $\mathcal{C}:=\mathcal{V} \backslash \{X,Y\}$. 

Studying the causal effect of $X$ on $Y$ means learning the (interventional) probability distribution, $\mathbb{P}(Y|do(X))$. We can write $\mathbb{P}(Y=y|do(X=x))$ as
\begin{align}
    \mathbb{P}(Y=y|do(X=x)) = \frac{\P(Y=y \cap do(X=x))}{\P(do(X=x))}
\end{align}
For estimating the above probability, we perform an RBD by forming blocks using covariates in $\mathcal{Z} \subseteq \mathcal{V} \backslash \{X,Y\}$, and similar to the estimation of $\E(Y|do(X))$, we define an estimate of $\mathbb{P}(Y=y|do(X=x))$ as
\begin{align}
    \hat{\P}_\mathcal{Z}&(Y=y|do(X=x)) := \sum_{\mathcal{Z} = z}\frac{\hat{\P}(Y=y \cap do(X=x) \cap \mathcal{Z} = z)}{\hat{\P}(do(X=x) \cap \mathcal{Z} = z))} \hat{\P}(\mathcal{Z}=z)
\end{align}
where $\hat{\P}(\cdot)$ are the sample relative frequencies.

It is desirable to select the $\mathcal{Z}$ such that $\Var(\hat{\P}_\mathcal{Z}(Y=y|do(X=x)))$ is minimized, i.e., maximum gain in precision. For ease of notation, we define $\Var(\hat{\P}_\mathcal{Z}(Y|do(X))):= \Var(\hat{\P}_\mathcal{Z}(Y=y|do(X=x)))$.

\begin{problem} \label{problem:optimal_blocking}  
Given a directed acyclic graph, $\mathcal{G}$, that depicts the causal relationships among some observed variables, $\mathcal{V}$, and unobserved variables, $\mathcal{U}$; treatment, $X\in\mathcal{V}$, and response, $Y\in\mathcal{V}$, obtain a smallest subset, $\mathcal{Z}^*$, of the set of observed covariates, $\mathcal{C}:=\mathcal{V} \backslash \{X,Y\}$, such that
\begin{align}
    \mathcal{Z}^* \in \argmin_{\mathcal{Z}}\Var(\hat{\P}_\mathcal{Z}(Y|do(X))).
\end{align}
\end{problem}

\begin{note}
\textup{
It is reasonable to assume that using some expert knowledge, we have access to the underlying causal graph $\mathcal{G}$ and do not know the strengths/weights of the causal connections depicted in this graph \citep{kyriacou2022using}. For instance, experts may tell us that alcohol consumption causes high blood pressure but may not have the knowledge of the magnitude of this causation. We can also construct the underlying causal graph using some observational data if it is available \citep{nordon2019building}.
}
\end{note}

\begin{note}
\textup{
While performing blocking for the minimization of the variance of the estimate of the causal effect, it is important to ensure that we are still able to accurately estimate the causal effect of interest. As introduced in \cref{sec:motivation}, we want the chosen set of covariates to be \textit{stable}, i.e. we exclude the covariates that one should never be used to create blocks. We formally motivate and define the notion of stability of the solution to \cref{problem:optimal_blocking} in \cref{sec:stable_blocking}. We also provide a method of achieving stability, i.e. finding and excluding the covariates that lead to inaccurate causal effect estimation. 
}
\end{note} 

\begin{note}
\textup{In experimental design, the term `block' refers to a subset of the experimental units that are homogenous with respect to some covariates while in causal graphical models, it refers to an obstruction in some causal path. For instance, for an experiment, a group of smokers can be a block while in causal graphical models, a common cause (confounder) blocks/obstructs a causal path.}
\end{note}

\section{Methodology} \label{sec:methodology}
In this section, we develop a methodology to find a solution to \cref{problem:optimal_blocking}. We first examine $\Var(\hat{\P}_\mathcal{Z}(Y|do(X)))$ as a function of $\mathcal{Z}$ to obviate some edges and nodes from the causal graph, $\mathcal{G}$. We next discuss some motivating examples and make key observations that lead to several lemmas for our main result. In the end, we provide our main result and develop an efficient algorithm for solving \cref{problem:optimal_blocking}.

\subsection{Examining $\Var(\hat{\P}_\mathcal{Z}(Y|do(X)))$ as a function of $\mathcal{Z}$}
In \cref{problem:optimal_blocking}, our interest is to minimize $\Var(\hat{\P}_\mathcal{Z}(Y|do(X)))$ as a function of $\mathcal{Z}\subseteq \mathcal{C}$. $\mathcal{Z}$ can affect $\Var(\hat{\P}_\mathcal{Z}(Y|do(X)))$ through the causal relationships from $\mathcal{Z}$ to $Y$, and from $\mathcal{Z}$ to $X$. Note that, $X$ is the treatment on which we are performing the intervention, i.e.,  we set the levels/values of $X$. This is equivalent to saying that $X$ is no longer affected by the rest of the variables. Therefore, $\mathcal{Z}$ also cannot affect $X$ when an intervention is performed on $X$. %instead its value is decided/set during the experiment. %In terms of the causal graph, $\mathcal{G}$, it is the same as deleting all arrows coming into $X$. 
Therefore, $\mathcal{Z}$ can affect $\Var(\hat{\P}_\mathcal{Z}(Y|do(X)))$ only through the causal relationships from $\mathcal{Z}$ to $Y$. Therefore, a smallest subset of $\mathcal{C}$ that blocks all causal paths from $\mathcal{C}$ to $Y$ gives a solution to \ref{problem:optimal_blocking}.

Denote the subgraph of $\mathcal{G}$ where all edges coming into $X$ were deleted as $\mathcal{G}_{\widetilde{X}}$.
\begin{lemma} \label{lemma:Gxbar}
For finding a solution to \cref{problem:optimal_blocking}, it is sufficient to work with the subgraph $\mathcal{G}_{\widetilde{X}}$.
\end{lemma}
\begin{proof}
The proof follows from the fact that when an intervention is performed on $X$ then it is no longer affected by the rest of the variables. In terms of the causal graph, $\mathcal{G}$, it is the same as deleting all edges coming into $X$. 
\end{proof}

\begin{definition}
$\mathcal{S},\mathcal{T}\subseteq \mathcal{C}$ are called equivalent sets (denoted as $\mathcal{S} \equiv \mathcal{T}$), if  $\Var(\hat{\P}_\mathcal{S}(Y|do(X))) = \Var(\hat{\P}_\mathcal{T}(Y|do(X)))$.
\end{definition}

\begin{lemma} \label{lemma:D0}
If $\mathcal{F}$ is the set of all covariates that do not have causal paths to $Y$, then $\mathcal{C}$ and $\mathcal{C}\backslash \mathcal{F}$ are equivalent sets.
\end{lemma}
\begin{proof}
The proof follows from the fact that a set of covariates can affect $\Var(\hat{\P}_\mathcal{Z}(Y|do(X)))$ only through the causal relationships from itself to $Y$. 
\end{proof}
Following \cref{lemma:Gxbar}, we restrict ourselves to the subgraph $\mathcal{G}_{\widetilde{X}}$. Following \cref{lemma:D0}, we further delete the covariates that do not have any causal paths to $Y$. Denote this new subgraph as $\mathcal{G}'_{\widetilde{X}}$ and $\mathcal{C}'=\mathcal{C}\backslash\mathcal{F}$. Therefore, a smallest subset of $\mathcal{C}$ that blocks all causal paths from $\mathcal{C}'$ to $Y$ in $\mathcal{G}'_{\widetilde{X}}$ gives a solution to \ref{problem:optimal_blocking}.

%As discussed earlier, we are interested in performing a randomized experiment to study the causal effect of interest $X\rightarrow Y$ where we are randomizing the treatment $(X)$. In terms of the causal graph, randomizing $X$ is the same as making interventions to set levels/values of $X$. This means that $X$ is not affected by any variable in the graph, instead, its value is decided/set during the experiment. Therefore, we can \textit{ignore all the arrows coming into $X$} and work with the subgraph of $G$ where all arrows coming into $X$ are ignored. We denote such subgraph as $G_{\widetilde{X}}$. Therefore, for the examples, we work with causal graphs with no arrows coming to $X$.

\subsection{Motivating Examples and Key Observations}

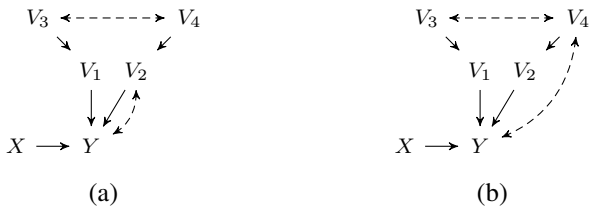
\begin{figure}[thb]
\centering
\begin{minipage}{.30\textwidth}
\centering
\begin{tikzpicture}[->,>=stealth',shorten >=1pt,auto,node distance=1cm,thin,main node/.style={font=\sffamily\small\bfseries}]
  \title{Figure 1}
  \node[main node] (X)  [] {\scriptsize $X$};
  \node[main node] (Y)  [right of=X] {\scriptsize $Y$};
  \node[main node] (Z1) [above of=Y] {\scriptsize $V_1$};
  \node[main node] (Z2) [right of=Z1, node distance=.6cm] {\scriptsize $V_2$};
  \node[main node] (Z3) [above left of=Z1] {\scriptsize $V_3$};
  \node[main node] (Z4) [above right of=Z2] {\scriptsize $V_4$};

  \begin{scope}
  \draw[->] (X) to node {} (Y);
  \draw[->] (Z1) to node {} (Y);
  \draw[->] (Z2) to node {} (Y);
  \draw[->] (Z3) to node {} (Z1);
  \draw[->] (Z4) to node {} (Z2);
  \draw[<->, densely dashed] (Z3) to node {} (Z4);
  \draw[<->, densely dashed, bend left] (Z2) to node {} (Y);
  \end{scope}
\end{tikzpicture}
\subcaption{}\label{fig:semi_markovian_model_Ya}
\end{minipage}\hspace{.5cm}
\begin{minipage}{.30\textwidth}
\centering
\begin{tikzpicture}[->,>=stealth',shorten >=1pt,auto,node distance=1cm,thin,main node/.style={font=\sffamily\small\bfseries}]
  \title{Figure 1}
  \node[main node] (X)  [] {\scriptsize $X$};
  \node[main node] (Y)  [right of=X] {\scriptsize $Y$};
  \node[main node] (Z1) [above of=Y] {\scriptsize $V_1$};
  \node[main node] (Z2) [right of=Z1, node distance=.6cm] {\scriptsize $V_2$};
  \node[main node] (Z3) [above left of=Z1] {\scriptsize $V_3$};
  \node[main node] (Z4) [above right of=Z2] {\scriptsize $V_4$};

  \begin{scope}
  \draw[->] (X) to node {} (Y);
  \draw[->] (Z1) to node {} (Y);
  \draw[->] (Z2) to node {} (Y);
  \draw[->] (Z3) to node {} (Z1);
  \draw[->] (Z4) to node {} (Z2);
  \draw[<->, densely dashed] (Z3) to node {} (Z4);
  \draw[<->, densely dashed, bend left] (Z4) to node {} (Y);
  \end{scope}
\end{tikzpicture}
\subcaption{}\label{fig:semi_markovian_model_Yb}
\end{minipage}\hspace{.5cm}
\caption{Example semi-Markovian model with latent variables involving $\mathcal{C}'$ and $Y$.}\label{fig:semi_markovian_model_Y}
\end{figure}

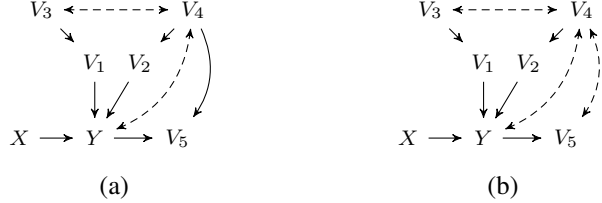
\begin{figure}[th]
\centering
\begin{minipage}{.30\textwidth}
\centering
\begin{tikzpicture}[->,>=stealth',shorten >=1pt,auto,node distance=1cm,thin,main node/.style={font=\sffamily\small\bfseries}]
  \title{Figure 1}
  \node[main node] (X)  [] {\scriptsize $X$};
  \node[main node] (Y)  [right of=X] {\scriptsize $Y$};
  \node[main node] (Z1) [above of=Y] {\scriptsize $V_1$};
  \node[main node] (Z2) [right of=Z1, node distance=.6cm] {\scriptsize $V_2$};
  \node[main node] (Z3) [above left of=Z1] {\scriptsize $V_3$};
  \node[main node] (Z5) [right of=Y, node distance=1.1cm] {\scriptsize $V_5$};
  \node[main node] (Z4) [above right of=Z2] {\scriptsize $V_4$};

  \begin{scope}
  \draw[->] (X) to node {} (Y);
  \draw[->] (Z1) to node {} (Y);
  \draw[->] (Z2) to node {} (Y);
  \draw[->] (Z3) to node {} (Z1);
  \draw[->] (Z4) to node {} (Z2);
  \draw[->] (Y) to node {} (Z5);
  \draw[->, bend left] (Z4) to node {} (Z5);
  \draw[<->, densely dashed] (Z3) to node {} (Z4);
  \draw[<->, densely dashed, bend left] (Z4) to node {} (Y);
  \end{scope}
\end{tikzpicture}
\subcaption{}\label{fig:semi_markovian_model_desa}
\end{minipage}\hspace{.5cm}
\begin{minipage}{.30\textwidth}
\centering
\begin{tikzpicture}[->,>=stealth',shorten >=1pt,auto,node distance=1cm,thin,main node/.style={font=\sffamily\small\bfseries}]
  \title{Figure 1}
  \node[main node] (X)  [] {\scriptsize $X$};
  \node[main node] (Y)  [right of=X] {\scriptsize $Y$};
  \node[main node] (Z1) [above of=Y] {\scriptsize $V_1$};
  \node[main node] (Z2) [right of=Z1, node distance=.6cm] {\scriptsize $V_2$};
  \node[main node] (Z3) [above left of=Z1] {\scriptsize $V_3$};
  \node[main node] (Z5) [ right of=Y, node distance=1.1cm] {\scriptsize $V_5$};
  \node[main node] (Z4) [above right of=Z2] {\scriptsize $V_4$};

  \begin{scope}
  \draw[->] (X) to node {} (Y);
  \draw[->] (Z1) to node {} (Y);
  \draw[->] (Z2) to node {} (Y);
  \draw[->] (Z3) to node {} (Z1);
  \draw[->] (Z4) to node {} (Z2);
  \draw[->] (Y) to node {} (Z5);
  \draw[<->, densely dashed, bend left] (Z4) to node {} (Z5);
  \draw[<->, densely dashed] (Z3) to node {} (Z4);
  \draw[<->, densely dashed, bend left] (Z4) to node {} (Y);
  \end{scope}
\end{tikzpicture}
\subcaption{}\label{fig:semi_markovian_model_desb}
\end{minipage}
\caption{Example Semi-Markovian Models with latent variables involving $\mathcal{C}'$ and $Y$, and $Y$ having descendants.}\label{fig:semi_markovian_model_des}
\end{figure}

All causal graphs in \cref{fig:markovian_model,fig:semi_markovian_model_C,fig:semi_markovian_model_Y,fig:semi_markovian_model_des,fig:semi_markovian_model_postx} represent $\mathcal{G}'_{\widetilde{X}}$. Denote the set of parents of $W$ in $\mathcal{G}'_{\widetilde{X}}$ as $\mathcal{P}(W)$.

\begin{note}
\textup{In all the graphs provided in this paper, a bi-directed edge between any two observed variables is the same as those two variables having one or more unobserved ancestors. For example $V_3 \dashleftarrow \dashrightarrow V_4$ is the same as $V_3 \dashleftarrow U \dashrightarrow V_4$ for some unobserved variable, $U$.
}
\end{note}

\begin{example} \label{example:1} 
\textup{
For \cref{fig:markovian_model,fig:semi_markovian_model_C}, $\mathcal{P}(Y)\backslash\{X\}=\{V_1,V_2\}$ is sufficient to block all causal paths from $\mathcal{C}'=\{V_1,V_2,V_3,V_4\}$ to $Y$. Therefore, $\mathcal{Z}^* = \mathcal{P}(Y)\backslash\{X\}$.}
\end{example}

\begin{figure}[th]
\centering
\begin{minipage}{.40\textwidth}
\centering
\begin{tikzpicture}[->,>=stealth',shorten >=1pt,auto,node distance=1cm,thin,main node/.style={font=\sffamily\small\bfseries}]
  \title{Figure 1}
  \node[main node] (X)  [] {\scriptsize $X$};
  \node[main node] (Y)  [right of=X] {\scriptsize $Y$};
  \node[main node] (Z1) [above of=Y] {\scriptsize $V_1$};
  \node[main node] (Z2) [right of=Z1, node distance=.7cm] {\scriptsize $V_2$};
  \node[main node] (Z3) [above left of=Z1] {\scriptsize $V_3$};
  \node[main node] (Z4) [above right of=Z2] {\scriptsize $V_4$};
  
  \begin{scope}
  \draw[->] (X) to node {} (Y);
  \draw[->] (Z1) to node {} (Y);
  \draw[->] (Z2) to node {} (Y);
  \draw[->] (Z3) to node {} (Z1);
  \draw[->] (Z4) to node {} (Z2);
  \end{scope}
\end{tikzpicture}
\caption{Example Markovian model (no latent structures).} \label{fig:markovian_model}
\end{minipage}\hspace{.5cm}
\begin{minipage}{.40\textwidth}
\centering
\begin{tikzpicture}[->,>=stealth',shorten >=1pt,auto,node distance=1cm,thin,main node/.style={font=\sffamily\small\bfseries}]
  \title{Figure 1}
  \node[main node] (X)  [] {\scriptsize $X$};
  \node[main node] (Y)  [right of=X] {\scriptsize  $Y$};
  \node[main node] (Z1) [above of=Y] {\scriptsize $V_1$};
  \node[main node] (Z2) [right of=Z1, node distance=.7cm] {\scriptsize $V_2$};
  \node[main node] (Z3) [above left of=Z1] {\scriptsize $V_3$};
  \node[main node] (Z4) [above right of=Z2] {\scriptsize $V_4$};

  \begin{scope}
  \draw[->] (X) to node {} (Y);
  \draw[->] (Z1) to node {} (Y);
  \draw[->] (Z2) to node {} (Y);
  \draw[->] (Z3) to node {} (Z1);
  \draw[->] (Z4) to node {} (Z2);
  \draw[<->, densely dashed] (Z3) to node {} (Z4);
  \end{scope}
\end{tikzpicture}
\caption{Example semi-Markovian model with latent variables involving only $\mathcal{C}'$.} \label{fig:semi_markovian_model_C}
\end{minipage}
\end{figure}

\begin{example} \label{example:2} 
\textup{
For \cref{fig:semi_markovian_model_Ya,fig:semi_markovian_model_Yb}, $\mathcal{P}(Y)\backslash\{X\}=\{V_1,V_2\}$ is not sufficient to block all causal paths from $\mathcal{C}'=\{V_1,V_2,V_3,V_4\}$ to $Y$. In \cref{fig:semi_markovian_model_Ya}, the reason is that blocking using $V_2$ opens the path $V_4 \rightarrow V_2 \dashleftarrow \dashrightarrow Y$ for $V_4$ to cause variability in $Y$. In \cref{fig:semi_markovian_model_Yb}, the reason is that the latent structure $Y \dashleftarrow \dashrightarrow V_4$ enables $V_4$ to cause $Y$ even when the path $V_4\rightarrow V_2\rightarrow Y$ is blocked. For both \cref{fig:semi_markovian_model_Ya,fig:semi_markovian_model_Yb}, $\mathcal{Z}^*=(\mathcal{P}(Y)\backslash\{X\}) \cup \{V_4\}$.
}
\end{example}

\begin{definition} \label{def:collider} \citep{pearl2009causality,rohrer2018thinking}
A variable is a collider or an inverted fork when it is causally influenced by two or more variables.
\end{definition}

\begin{note}
\textup{
If we condition/block using a collider (and/or it's descendants) then the common causes of the collider become statistically dependent \citep{berkson1946limitations,cole2010illustrating}. For example, if alcohol and smoking are common causes of high blood pressure then it is not necessary that these common causes are correlated in the population at large. However, if we restrict our attention to (condition on) those with high blood pressure then we introduce a non-causal dependence between alcohol and smoking. \footnote{The reader may refer to the tutorials \href{https://www.col-ex.org/posts/conditioning-causal-graphs/\#d-separation-and-d-connection-along-a-path}{Collectively Exhaustive} and \href{https://donskerclass.github.io/CausalEconometrics/DAGs.html}{Causal Graphs by David Childers} for understanding the basic concepts related to causal graphs.}
}
\end{note} 

\begin{example} \label{example:3} 
\textup{
For \cref{fig:semi_markovian_model_des}, despite $Y$ having descendants, $(\mathcal{P}(Y)\backslash\{X\}) \cup \{V_4\}$ blocks all causal paths from $\mathcal{C}'$ to $Y$. This is because descendants of $Y$ can never be a cause of variability in $Y$. In \cref{fig:semi_markovian_model_desa,fig:semi_markovian_model_desb}, $V_5$ (a descendant of $Y$) cannot cause $Y$ through the edge $Y \rightarrow V_5$. Moreover, $V_4$ (an ancestor of $Y$) cannot cause $Y$ through $V_5$ because $V_5$ is a collider on the path $V_4\rightarrow V_5 \leftarrow Y$.
}
\end{example}

Denote the set of ancestors (including itself) of $W$ in $\mathcal{G}'_{\widetilde{X}}$ as $\mathcal{A}(W)$. Denote the subgraph of $\mathcal{G}'_{\widetilde{X}}$ restricted to $\mathcal{A}(Y)$ as $\mathcal{G}'_{\widetilde{X},\mathcal{A}(Y)}$. Based on the insights from \cref{example:1,example:2,example:3}, we write \cref{lemma:GxbarA}.
\begin{lemma}\label{lemma:GxbarA}
For finding a solution to \cref{problem:optimal_blocking}, it is sufficient to work with the subgraph $\mathcal{G}'_{\widetilde{X},\mathcal{A}(Y)}$.
\end{lemma}

\begin{proof}
The proof follows from the fact the graph under consideration is acyclic. Hence, the descendants of a node can never be its causes even in the presence of latent structures.
\end{proof}

\subsection{Solution to Problem \ref{problem:optimal_blocking}}

We now combine the insights from \cref{lemma:Gxbar,lemma:D0,lemma:GxbarA} to provide a solution to \cref{problem:optimal_blocking}.

\begin{theorem} \label{theorem:sol_to_optimal_blocking}
The smallest $\mathcal{Z}$ such that $Y \perp (\mathcal{A}(Y)\backslash \{X,Y\})\backslash\mathcal{Z}|\mathcal{Z}$ in $\mathcal{G}'_{\widetilde{X},\mathcal{A}(Y)}$ is a solution to \cref{problem:optimal_blocking}.
\end{theorem}
\begin{proof}
The proof follows from the fact that $\mathcal{Z}$ such that $Y \perp (\mathcal{A}(Y)\backslash \{X,Y\})\backslash\mathcal{Z}|\mathcal{Z}$ in $\mathcal{G}'_{\widetilde{X},\mathcal{A}(Y)}$ blocks all causal paths from $\mathcal{C}'$ to $Y$. %following \cref{lemma:Gxbar,lemma:D0,lemma:GxbarA}.
\end{proof}

\cref{theorem:sol_to_optimal_blocking} provides a sufficient condition for a set $\mathcal{Z}$ to be a solution to \cref{problem:optimal_blocking}. We next provide a method to construct a set $\mathcal{Z}$ that satisfies this sufficient condition.

Let a path composed entirely of bi-directed edges be called a \textit{bi-directed path}.
\begin{definition}\label{def:c_component} \citep{tian2002general}. 
For a graph $\mathcal{G}$, the set of observed variables $\mathcal{V}$, can be \textit{partitioned} into disjoint groups by assigning two variables to the same group if and only if they are connected by a bi-directed path. Assume that $V$ is therefore partitioned into $k$ groups $\mathcal{S}_1, \dots, \mathcal{S}_k$. We will call each $\mathcal{S}_j; j=1, \dots, k,$ a \textit{$c$-component} of $V$ in $\mathcal{G}$ or a $c$-component (abbreviating confounded component) of $\mathcal{G}$. The $\mathcal{S}_j$ such that $W\in S_j$ is called the \textit{$c$-component of $W$} in $\mathcal{G}$ and is denoted as $C_{W,\mathcal{G}}$. As $\{\mathcal{S}_1, \dots, \mathcal{S}_k\}$ is a partition of $\mathcal{V}$, $c$-component of a variable always exists and is unique.
\end{definition}

In \cref{example:2}, we observed that due to the presence of latent structures involving $Y$, $\mathcal{P}(Y)$ was not sufficient to block all causal paths from $\mathcal{C}'$ to $Y$. However the parents of $c$-component of $Y$ in $\mathcal{G}'_{\widetilde{X},\mathcal{A}(Y)}$ would be sufficient for that purpose. 

\begin{theorem} \label{theorem:sol_to_optimal_blocking_construction}
The smallest $\mathcal{Z}$ such that $Y \perp (\mathcal{A}(Y)\backslash \{X,Y\})\backslash\mathcal{Z}|\mathcal{Z}$ is  the set of parents of c-component of $Y$ in $\mathcal{G}'_{\widetilde{X},\mathcal{A}(Y)}$ excluding $X$ denoted as $\mathcal{P}(C_{Y,\mathcal{G}_{\widetilde{X},\mathcal{A}(Y)}})\backslash\{X\}$.
\end{theorem} 
\begin{proof}
The proof follows from Corollary 1 \citep{tian2002studies}, which states that a node is independent of its ancestors excluding the parents of its $c$-component given the parents of its $c$-component.
\end{proof} 

\begin{note} 
\textup{
In a Markovian model there are no bi-directed paths, Therefore, the elementary partition of $\mathcal{A}(Y)$ is the $c$-component of $\mathcal{G}'_{\widetilde{X},\mathcal{A}(Y)}$. Therefore, $C_Y = \{Y\}$, and $\mathcal{P}(C_{Y,\mathcal{G}_{\widetilde{X},\mathcal{A}(Y)}}) \backslash\{X\} = \mathcal{P}(Y) \backslash\{X\}$ in $\mathcal{G}'_{\widetilde{X},\mathcal{A}(Y)}$. This matches with our insight in \cref{example:1}, where we observed that $Y \perp (\mathcal{A}(Y)\backslash\{X,Y\})\backslash(\mathcal{P}(Y)\backslash\{X\})|(\mathcal{P}(Y)\backslash\{X\})$.
}
\end{note}

\subsection{Stability of the Solution to Problem \ref{problem:optimal_blocking}} \label{sec:stable_blocking}

We next discuss some concerns with using $\mathcal{Z}^*$ (a solution to \cref{problem:optimal_blocking}) as a blocking set. We provide a method to address those concerns. 
\begin{definition} \label{def:ptay}
Post-treatment ancestors of  (excluding $X$) of the response are those ancestors of  (excluding $X$) of the response who are also the descendants of the treatment.
\end{definition}

Denote the post-treatment ancestors of the response while studying the causal effect $Y|do(X)$ as $\mathcal{M}(Y|do(X))$, and the set of descendants (including itself) of $W$ in $\mathcal{G}'_{\widetilde{X},\mathcal{A}(Y)}$ as $\mathcal{D}(W)$.

%Post-treatment ancestors (excluding itself) of the response are those ancestors (excluding itself) of the response who are also the descendants of the treatment. 
%Denote the set of ancestors of $W$ in the subgraph of $\mathcal{G}'_{\widetilde{X},\mathcal{A}(Y)}$ where all edges leaving $X$ as $\mathcal{A}$.
\begin{lemma} \label{lemma:M}
$\mathcal{M}(Y|do(X)) = (\mathcal{D}(X) \cap \mathcal{A}(Y))\backslash \{X,Y\}$.
\end{lemma}
\begin{proof}
The proof follows directly from \cref{def:ptay}. %i.e. post-treatment ancestors (excluding itself) of $Y$.
\end{proof}

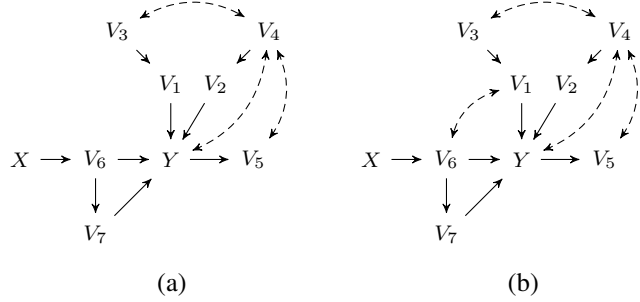
\begin{figure}[t]
\centering
\begin{minipage}{.30\textwidth}
\begin{tikzpicture}[->,>=stealth',shorten >=1pt,auto,node distance=1cm,thin,main node/.style={font=\sffamily\small\bfseries}]
  \title{Figure 1}
  \node[main node] (X)  [] {\scriptsize $X$};
  \node[main node] (Z6) [right of=X] {\scriptsize $V_6$};
  \node[main node] (Y)  [right of=Z6] {\scriptsize $Y$};
  \node[main node] (Z1) [above of=Y] {\scriptsize $V_1$};
  \node[main node] (Z2) [right of=Z1, node distance=.6cm] {\scriptsize $V_2$};
  \node[main node] (Z3) [above left of=Z1] {\scriptsize $V_3$};
  \node[main node] (Z4) [above right of=Z2] {\scriptsize $V_4$};
  \node[main node] (Z5) [right of=Y, node distance=1.1cm] {\scriptsize $V_5$};
  \node[main node] (Z7) [below of=Z6] {\scriptsize $V_7$};

  \begin{scope}
  \draw[->] (X) to node {} (Z6);
  \draw[->] (Y) to node {} (Z5);
  \draw[->] (Z1) to node {} (Y);
  \draw[->] (Z2) to node {} (Y);
  \draw[->] (Z3) to node {} (Z1);
  \draw[->] (Z6) to node {} (Z7);
  \draw[->] (Z6) to node {} (Y);  
  \draw[->] (Z4) to node {} (Z2); 
  \draw[->] (Z7) to node {} (Y); 
  \draw[<->, densely dashed, bend left] (Z4) to node {} (Z5);
  \draw[<->, densely dashed, bend left] (Z3) to node {} (Z4);
  \draw[<->, densely dashed, bend left] (Z4) to node {} (Y);
  \end{scope}
\end{tikzpicture}
\subcaption{}\label{fig:semi_markovian_model_postxa}
\end{minipage}
\begin{minipage}{.30\textwidth}
\begin{tikzpicture}[->,>=stealth',shorten >=1pt,auto,node distance=1cm,thin,main node/.style={font=\sffamily\small\bfseries}]
  \title{Figure 1}
  \node[main node] (X)  [] {\scriptsize $X$};
  \node[main node] (Z6) [right of=X] {\scriptsize $V_6$};
  \node[main node] (Y)  [right of=Z6] {\scriptsize $Y$};
  \node[main node] (Z1) [above of=Y] {\scriptsize $V_1$};
  \node[main node] (Z2) [right of=Z1, node distance=.6cm] {\scriptsize $V_2$};
  \node[main node] (Z3) [above left of=Z1] {\scriptsize $V_3$};
  \node[main node] (Z4) [above right of=Z2] {\scriptsize $V_4$};
  \node[main node] (Z5) [right of=Y, node distance=1.1cm] {\scriptsize $V_5$};
  \node[main node] (Z7) [below of=Z6] {\scriptsize $V_7$};

  \begin{scope}
  \draw[->] (X) to node {} (Z6);
  \draw[->] (Y) to node {} (Z5);
  \draw[->] (Z1) to node {} (Y);
  \draw[->] (Z2) to node {} (Y);
  \draw[->] (Z3) to node {} (Z1);
  \draw[->] (Z6) to node {} (Z7);
  \draw[->] (Z6) to node {} (Y);  
  \draw[->] (Z4) to node {} (Z2); 
  \draw[->] (Z7) to node {} (Y); 
  \draw[<->, densely dashed, bend right] (Z1) to node {} (Z6);
  \draw[<->, densely dashed, bend left] (Z4) to node {} (Z5);
  \draw[<->, densely dashed, bend left] (Z3) to node {} (Z4);
  \draw[<->, densely dashed, bend left] (Z4) to node {} (Y);
  \end{scope}
\end{tikzpicture}
\subcaption{}\label{fig:semi_markovian_model_postxb}
\end{minipage}
\caption{Example semi-Markovian models with post-treatment ancestors of $Y$.}\label{fig:semi_markovian_model_postx}
\end{figure}

\begin{example} \label{example:4} 
\textup{In \cref{fig:semi_markovian_model_postxa,fig:semi_markovian_model_postxb}, there exist post-treatment ancestors (excluding itself) of the response. Blocks created using these covariates are not well-defined. For instance, we are interested in studying the causal effect of a drug on blood pressure. Suppose anxiety level mediates the effect of drug on blood pressure, i.e., $\text{drug}\rightarrow\text{anxiety}\rightarrow \text{blood pressure}$. If we create blocks using anxiety then the blocks will change (i.e., become unstable) during the experiment because change in the drug level will cause the anxiety level to change. Therefore, it is reasonable not to create blocks using these covariates. With this insight, for \cref{fig:semi_markovian_model_postxa}, $(\mathcal{P}(Y)\backslash\{X\})\cup\{V_4\}$ blocks all (pre-treatment) causal paths from $\mathcal{C}'$ to $Y$.  However, we cannot simply delete the post-treatment ancestors (excluding itself) of the response from the subgraph $\mathcal{G}'_{\widetilde{X},\mathcal{A}(Y)}$ of interest. This is because the presence of latent structures between pre- and post-treatment ancestors of the response affects blocking set of interest. For instance, in \cref{fig:semi_markovian_model_postxb}, $V_1$ is a collider on the path $V_3\rightarrow V_1\leftarrow V_6$. Hence, keeping $V_1$ in the blocking set will open this path for $V_3$ to cause variability in $Y$. Therefore, for \cref{fig:semi_markovian_model_postxb}, we need $(\mathcal{P}(Y)\backslash\{X\})\cup\{V_3,V_4\}$ to block all (pre-treatment) causal paths from $\mathcal{C}'$ to $Y$.
}
\end{example}

% To formalize the insight from Example 4, we refer to the following definition.
% \begin{definition} \cite{shpitser2012validity} A causal path from a variable $X \in \mathcal{X}$ to $Y \in \mathcal{Y}$ is called \textit{proper} if it doesn't intersect $\mathcal{X}$ except the endpoints. When $\mathcal{X}$ is a single variable then all causal paths are proper.
% \end{definition}

\begin{definition}
For studying the causal effect of $X$ on $Y$, a blocking set $\mathcal{Z}$, is said to be stable if $\mathcal{Z} \cap \mathcal{M}(Y|do(X)) = \phi$.
\end{definition}

\begin{theorem}\label{theorem:stable_blocking}
A stable solution to \cref{problem:optimal_blocking} is $\mathcal{Z}^* = (\mathcal{P}(C_{Y,\mathcal{G}_{\widetilde{X},\mathcal{A}(Y)}} \backslash\{X\}) \backslash \mathcal{M}(Y|do(X))$.
\end{theorem}
\begin{proof}
The proof follows from \cref{lemma:M}.
\end{proof}

%Therefore, we ignore all the variables falling on a proper causal path from $X$ to $Y$. Following \cite{van2014constructing}, the set of all such variables is $Dpcp(X,Y) = De((De_{\widetilde{X}}(X) \backslash X) \cap An_{\tilde{\tilde{X}}(Y))$ where $De(.)$ represents the set of descendants excluding the variable itself and $An_{\tilde{\tilde{X}}(Y)}$ is the ancestral graph of $Y$ where all edges leaving $X$ are removed.

%Note that by definition $Dpcp(X,Y)$ may include non-descendants of $Y$ which we have already argued to not take into consideration in Observation 2. Therefore, we can say that specifically we should not block using  $Dpcp(X,Y)$ in the graph $G_{An_{\widetilde{X}}(Y)}$ denoted as $Dpcp(X,Y)_{G_{An_{\widetilde{X}}(Y)}}$. Note that $Dpcp(X,Y)_{G_{An_{\widetilde{X}}(Y)}} = Dpcp(X,Y) \cap An_{\widetilde{X}}(Y)$.

\subsection{Final Algorithm}

We next provide steps to obtain a stable solution to \cref{problem:optimal_blocking} based on \cref{theorem:sol_to_optimal_blocking,theorem:sol_to_optimal_blocking_construction,theorem:stable_blocking}. First, we reduce $\mathcal{G}$ to $\mathcal{G}'_{\widetilde{X},\mathcal{A}(Y)}$. Next, we obtain the set $\mathcal{Z} = \mathcal{P}(C_{Y,\mathcal{G}_{\widetilde{X},\mathcal{A}(Y)}})\backslash\{X\}$ which is the smallest set such that $Y \perp (\mathcal{A}(Y)\backslash \{X,Y\})\backslash\mathcal{Z}|\mathcal{Z}$ in $\mathcal{G}'_{\widetilde{X},\mathcal{A}(Y)}$. Finally, we drop $\mathcal{M}(Y|do(X))$ from $\mathcal{Z}$ to get a stable solution to \cref{problem:optimal_blocking}.

Denote the edge $W_1\rightarrow W_2$ as \textit{directed-edge}$(W_1,W_2)$, a path consisting of directed edges from $W_1$ to $W_2$ as \textit{directed-path}$(W_1,W_2)$, and a path consisting of all bi-directed edges from $W_1$ to $W_2$ as \textit{bi-directed-path}$(W_1,W_2)$. Algorithm \ref{alg:causal-opt-blocking-set} outlines the psuedocode of the proposed algorithm to obtain a stable solution to \cref{problem:optimal_blocking}.

% \if 0
\begin{algorithm} [!htb]
	\caption{Stable-Causal-Blocking ($\mathcal{G}, X, Y$)}\label{alg:causal-opt-blocking-set}
	\begin{algorithmic}[1]
	    %\State \textbf{Input} {($\mathcal{G}, X, Y$)}.
        \For{$i$ in $1,\dots,|\mathcal{V}|$} \Comment{Reducing $\mathcal{G}$ to $\mathcal{G}_{\widetilde{X}}$} 
            \If{$\exists$ \textit{directed-edge}($V_i,X$)}
                \State delete \textit{directed-edge}($V_i,X$) from $\mathcal{G}$
            \EndIf
        \EndFor 
        \For{$i$ in $1,\dots,|\mathcal{V}|$} \Comment{Reducing $\mathcal{G}_{\widetilde{X}}$ to $\mathcal{G}'_{\widetilde{X}}$} 
            \If{$\nexists$ \textit{directed-path}($V_i,Y$)}
                \State delete $V_i$ from $\mathcal{G}$
            \EndIf
        \EndFor 
        \State $\mathcal{A}(Y) \leftarrow \{Y\}$
        \For{$i$ in $1,\dots,|\mathcal{V}|$} \Comment{Reducing $\mathcal{G}'_{\widetilde{X}}$ to $\mathcal{G}'_{\widetilde{X},\mathcal{A}(Y)}$} 
        \If{$\exists$ path($V_i,Y$)}
            \State $\mathcal{A}(Y) \leftarrow \mathcal{A}(Y) \cup \{V_i\}$
        \EndIf
        \EndFor
        %\State $\mathcal{A}(Y) = \mathcal{V}\backslash\mathcal{A}'(Y)$
        \State $\mathcal{S}_i = \{V_i\}, i =1,\dots,|\mathcal{V}|$ \Comment{Finding $C_{Y,\mathcal{G}_{\widetilde{X},\mathcal{A}(Y)}}$}
        \For{$i$ in $1,\dots,|\mathcal{V}|$}
            \For{$j$ in $1,\dots,i$}
            \If{$\exists$ \textit{bi-directed-path}($V_i$,$V_j$)}
                \State $\mathcal{S}_i \leftarrow \mathcal{S}_i \cup \mathcal{S}_j; \mathcal{S}_j \leftarrow \phi$
            \EndIf
            \EndFor
        \EndFor
        \State $i\leftarrow 1; \mathcal{S} \leftarrow \mathcal{S}_1$
        \While{$Y \notin \mathcal{S}$} 
            \State $i \leftarrow i+1; \mathcal{S} \leftarrow \mathcal{S}_i$
        \EndWhile
        \State $\mathcal{P}(\mathcal{S})=\phi$ \Comment{Finding $\mathcal{P}(C_{Y,\mathcal{G}_{\widetilde{X},\mathcal{A}(Y)}})$}
        \For{$i$ in $1,\dots,|\mathcal{V}|$} 
            \For{$W$ in $\mathcal{S}$}
                \If{$\exists$ \textit{directed-edge}($V_i,W$)}
                    \State $\mathcal{P}(\mathcal{S}) \leftarrow \mathcal{P}(\mathcal{S}) \cup \{V_i\}$
                \EndIf
            \EndFor
        \EndFor 
    	\State $\mathcal{D}(X) \leftarrow \{X\}$ \Comment{Finding $\mathcal{D}(X)$} 
        \For{$i$ in $1,\dots,|\mathcal{V}|$} 
        \If{$\exists$ \textit{directed-path}($X,V_i$)}
            \State $\mathcal{D}(X) \leftarrow \mathcal{D}(X) \cup \{V_i\}$
        \EndIf
        \EndFor  
        %\State $\mathcal{D}(X) = \mathcal{V}\backslash\mathcal{D}'(X)$
    	\State $\mathcal{Z} \leftarrow (\mathcal{P}(\mathcal{S})\backslash\{X\}) \backslash (\mathcal{D}(X) \cap \mathcal{A}(Y))\backslash \{X,Y\}$.
    	\State \textbf{return} $\mathcal{Z}$
	\end{algorithmic}
\end{algorithm}
% \fi

\subsection{Time Complexity of Algorithm \ref{alg:causal-opt-blocking-set}}
We next show that \cref{alg:causal-opt-blocking-set} is an efficient polynomial-time algorithm. Finding if there exists a path between any two nodes in a directed graph can be done using breath-first-search which has the worst-case time complexity of $O(|\mathcal{V}|^2)$. \cref{alg:causal-opt-blocking-set} first reduces $\mathcal{G}$ to $\mathcal{G}'_{\widetilde{X},\mathcal{A}(Y)}$ which involves finding variables with edges going into $X$ taking $O(|\mathcal{V}|)$ steps, finding variables with no path into $Y$ taking $O(|\mathcal{V}|^3)$ steps, and finding all ancestors of $Y$ taking $O(|\mathcal{V}|^3)$ steps. \cref{alg:causal-opt-blocking-set} next finds the $c$-component of $Y$ involves traversing all pairs of variables: $O(|\mathcal{V}|^2)$, and finding if there exists a bi-directed path between any pair of variables: $O(|\mathcal{V}|^2)$. Therefore, the total time at this step is $O(|\mathcal{V}|^4)$. Finally \cref{alg:causal-opt-blocking-set} finds $\mathcal{M}(Y|do(X))$ which involves finding descendants of $X$ taking $O(|\mathcal{V}|^3)$ steps, and finding ancestors of $Y$ taking $O(|\mathcal{V}|^3)$ steps. Therefore, the time complexity of \cref{alg:causal-opt-blocking-set} is $O(|\mathcal{V}|^4)$.

% \begin{enumerate}
%     \item finding variables with edges going into $X$: $O(|\mathcal{V}|)$, 
%     \item finding variables with no path into $Y$: $O(|\mathcal{V}|^3)$, and 
%     \item finding all ancestors of $Y$: $O(|\mathcal{V}|^3)$.
% \end{enumerate}

% \begin{enumerate}
%     \item finding descendants of $X$: $O(|\mathcal{V}|^3)$, and
%     \item finding ancestors of $Y$: $O(|\mathcal{V}|^3)$.
% \end{enumerate}

\subsection{Implementing Algorithm \ref{alg:causal-opt-blocking-set}} \label{sec:implementingthealgo}
\begin{figure}[h!]
\centering
\begin{tikzpicture}[->,>=stealth',shorten >=1pt,auto,node distance=1.5cm,thin,main node/.style={font=\sffamily\small\bfseries}]
  \title{Figure 1}
  \node[main node] (X) {\scriptsize Drug ($X$)};
  \node[main node] (Z5) [right of=X, node distance = 2cm] {\scriptsize Anxiety};
  \node[main node] (Y)  [right of=Z5, node distance = 3cm] {\scriptsize Blood Pressure ($Y$)};
  \node[main node] (Z1) [above of=Y] {\scriptsize Alcohol};
  \node[main node] (Z2) [right of=Z1, node distance = 3cm] {\scriptsize Cholesterol};
  \node[main node] (Z3) [above of=X] {\scriptsize Food-Habits};
  \node[main node] (Z4) [right of=Z2, node distance = 2cm] {\scriptsize Age};
  \node[main node] (Z6) [below of=Z5] {\scriptsize Sleep-Quality};
  \node[main node] (Z7) [below of=Y] {\scriptsize Palpitations};
  \node[main node] (Z8) [right of=Z7, node distance = 3cm] {\scriptsize Strenuous-Activity};
  \node[main node] (Z9) [left of=X, node distance = 2.5cm] {\scriptsize Blood-Sugar};
  
  \begin{scope}
  \draw[->] (X) to node {} (Z5);
  \draw[->] (Z5) to node {} (Y);
  \draw[->] (Z5) to node {} (Z6);
  \draw[->] (Y) to node {} (Z7);
  \draw[->] (Z1) to node {} (Y);
  \draw[->, bend right] (Z1) to node {} (X);
  \draw[->] (Z3) to node {} (X);
  \draw[->] (Z2) to node {} (Y);
  \draw[->, bend left] (Z3) to node {} (Z1);
  \draw[->] (Z4) to node {} (Z2);
  \draw[->] (Z8) to node {} (Z7);
  \draw[->] (Z9) to node {} (X);
  \draw[->] (Z6) to node {} (Y);
  \draw[->, bend left] (Z2) to node {} (Z7);
  \draw[<->, densely dashed, bend left] (Z2) to node {} (Y);
  \draw[<->, densely dashed, bend left] (Z5) to node {} (Z1);
  \draw[<->, densely dashed, bend left] (Z5) to node {} (Y);
  \end{scope}  
\end{tikzpicture}
\caption{The Original Graph $\mathcal{G}$.} \label{fig:final_implementation1G}
\end{figure}
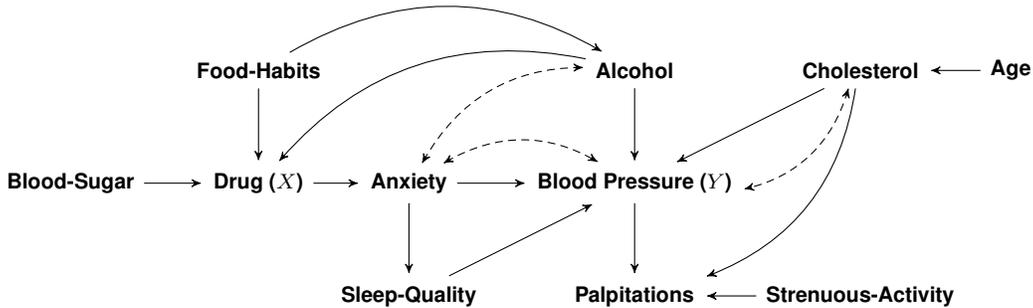

We demonstrate the implementation of \cref{alg:causal-opt-blocking-set} to obtain a stable solution to \cref{problem:optimal_blocking} for a general semi-Markovian causal graph, $\mathcal{G}$ given in \cref{fig:final_implementation1G}.

\if 0
\begin{figure}[h!]
\centering
\begin{tikzpicture}[->,>=stealth',shorten >=1pt,auto,node distance=1cm,thin,main node/.style={font=\sffamily\small\bfseries}]
  \title{Figure 1}
  \node[main node] (X) {\scriptsize Drug ($X$)};
  \node[main node] (Z5) [right of=X, node distance = 2cm] {\scriptsize Anxiety};
  \node[main node] (Y)  [right of=Z5, , node distance = 3cm] {\scriptsize Blood Pressure ($Y$)};
  \node[main node] (Z1) [above of=Y] {\scriptsize Alcohol};
  \node[main node] (Z2) [right of=Z1, node distance = 3cm] {\scriptsize Cholesterol};
  \node[main node] (Z3) [left of=Z1, node distance = 4cm] {\scriptsize Food-Habits};
  \node[main node] (Z4) [right of=Z2, node distance = 2cm] {\scriptsize Age};
  \node[main node] (Z6) [below of=Z5] {\scriptsize Sleep-Quality};
  
  \begin{scope}
  \draw[->] (X) to node {} (Z5);
  \draw[->] (Z5) to node {} (Y);
  \draw[->] (Z5) to node {} (Z6);
  \draw[->] (Z1) to node {} (Y);
  \draw[->] (Z2) to node {} (Y);
  \draw[->, bend left] (Z3) to node {} (Z1);
  \draw[->] (Z4) to node {} (Z2);
  \draw[->] (Z6) to node {} (Y);
  \draw[<->, densely dashed, bend left] (Z2) to node {} (Y);
  \draw[<->, densely dashed, bend left] (Z5) to node {} (Z1);
  \draw[<->, densely dashed, bend left] (Z5) to node {} (Y);
  \end{scope}  
\end{tikzpicture}
\caption{The Graph $\mathcal{G}_{\widetilde{X},\mathcal{A}(Y)}$.}\label{fig:final_implementation1Gnew}
\end{figure}
\fi

We first obtain the graph $G_{An_{\widetilde{X}}(Y)}$. %as displayed in \cref{fig:final_implementation1Gnew}. 
We next obtain the $c$-components (following \cref{def:c_component}) of $\mathcal{G}_{\widetilde{X},\mathcal{A}(Y)}$ are $\{X\}, \{\text{Alcohol}, \text{Anxiety}, Y, \text{Cholesterol}\}, \{\text{Food-Habits}\}, \{\text{Age}\}$, and the $c$-component of $Y$ is $\{\text{Alcohol},\text{Anxiety},Y,\text{Cholesterol}\}$. We next obtain the set of parents of the $c$-component of $Y$ as %$\mathcal{P}(C_{Y,\mathcal{G}_{\widetilde{X},\mathcal{A}(Y)}})\backslash\{X\} =
$\{X, \text{Alcohol}, \text{Cholesterol}, \text{Food-Habits}, \text{Age}, \text{Anxiety}\}$. The set $\mathcal{D}(X) \cap \mathcal{A}(Y)$ is $\{X,Y,\text{Anxiety},\text{Sleep-Quality}\}$. Finally, we obtain the final set as:
$$(\{X, \text{Alcohol}, \text{Cholesterol}, \text{Food-Habits}, \text{Age}, \text{Anxiety}\}\backslash\{X\})\backslash(\{X,Y,\text{Anxiety},\text{Sleep-Quality}\}\backslash\{X,Y\})$$ 
$$= \{\text{Alcohol}, \text{Cholesterol}, \text{Food-Habits}, \text{Age}\}.$$
\section{Experimental Evaluations} \label{sec:experiments}

In this section, we provide experimental evaluations to compare our method with other choices of sets of covariates to create blocks. For the causal graph, $\mathcal{G}$ provided in \cref{fig:final_implementation1G}, we are interested in studying the causal effect of Drug ($X$) on Blood-Pressure ($Y$). 

We considered a simple setting where all variables under consideration are binary variables taking values in \{0,1\}. We used a simple joint probability distribution that factorizes according to the given causal graph. We then performed simulated randomized experiments for different choices of sets of covariates to create blocks. The estimated causal effect and the variability in the response for different choices of blocking sets are provided in \cref{tab:causal_effect}. We observe that the variability in the response is the smallest for the choice of covariates obtained using the algorithm proposed in this paper. For details of the experimental evaluations, refer to \cref{app:exp_details}.

\begin{table*}[t] 
    \centering 
    \caption{Causal effect estimate and variability in the response.}  \label{tab:causal_effect}
	\begin{tabular}{l r r}
		\hline \hline
		  Blocking set & Causal effect & Variance \\
            \hline
            $\phi$, i.e. no blocking & 0.6910 & 0.2126 \\
            \{Food Habits\} & 0.6950 & 0.2098 \\
            \{Food Habits, Alcohol\} & 0.6960 & 0.2028 \\    
            \{Food Habits, Alcohol, Cholesterol\} & 0.6760 & 0.2001 \\     
            \{Food Habits, Alcohol, Cholesterol, Age\} & 0.7150 & 0.1754\\
	    \hline
	    \hline
	\end{tabular}
\end{table*}
\section{Conclusion and Future Work} \label{sec:conclusion}
We investigated the problem of finding a stable set of covariates to be used for forming blocks, that minimizes the variance of the causal effect estimates. By leveraging the underlying causal graph, we provided an efficient algorithm to obtain such a set for a general semi-Markovian causal model. 

In the future, we are interested in extending our work towards further pruning the stable causal blocking set obtained in this paper for the reasons of economic cost. If the experimenter has a constraint on the cardinality (size) of the blocking set due to budget limitations, i.e. $|\mathcal{Z}| \le k_1$ where $k_1$ is specified by the experimenter. Furthermore, if blocking variables have different economic costs associated with them, for example, getting data on cholesterol level may be costlier (or difficult) as compared to the age of the patient, then we can introduce a knapsack budget constraint. Let $e(Z)$ be the cost associated with $Z\in \mathcal{Z}$ then the knapsack constraint would be $\sum_{Z\in\mathcal{Z}}e(Z) \le k_2$ where $k_2$ is specified by the experimenter. We are interested in solving the problem of finding stable solutions to \cref{problem:optimal_blocking} with additional cardinality or knapsack constraint on the feasible solutions. Furthermore, we are interested in extending our work for the cases when the treatment and response variables are sets of variables. For the future direction to use the proposed algorithm for different real-world applications, refer to \cref{app:future}. \\

%\acks{The author thanks the anonymous reviewers, Dr. Christopher J. Quinn from Iowa State University, and Dr. Vaneet Aggarwal from Purdue University, for their valuable suggestions.}

\clearpage
\bibliography{refs.bib}

\clearpage
\appendix
\section{Expectation and Variance of Causal Effect without and with Blocking}

In this section, we calculate the expectation and variance of the causal effect  without blocking (defined in \eqref{eq:beta}) and  with blocking (defined in \eqref{eq:beta_z}).

\subsection{Expectation and Variance of $\hat{\beta}$} \label{app:proofs_hat_beta}

We calculate the expectation of $\hat{\beta}$ as
\begin{align}
\E(\hat{\beta}) &= \E(\bar{Y}(1)) - \E(\bar{Y}(0)). \label{eq:ebeta}
\end{align}
In general, due to non-homogeneous experimental units, $\E(\bar{Y}(x)) \ne \E(Y(x)), x=0,1$. Let $\mathcal{Z}$ be the covariates causing non-homogeneity, rewrite \eqref{eq:ebeta} as
\begin{align}
\E(\hat{\beta}) &= \E_\mathcal{Z}\left(\E(\bar{Y}(1)|\mathcal{Z})\right) - \E_\mathcal{Z}\left(\E(\bar{Y}(0)|\mathcal{Z})\right).
\end{align}
For fixed covariates the experimental units are identical. Therefore, $\E(\bar{Y}(x)|\mathcal{Z}) = \E(Y(1)|\mathcal{Z}), x=0,1$. Thus,
\begin{align}
\E(\hat{\beta}) &= \E_\mathcal{Z}\left(\E({Y}(x)|\mathcal{Z})\right) - \E_\mathcal{Z}\left(\E({Y}(0)|\mathcal{Z})\right),\\
&= \E(Y(1)) -  \E(Y(0)) = \beta.
\end{align}
%Therefore, the unbiasedness of $\hat{\beta}$ for $\beta$.
We next calculate the variance of $\hat{\beta}$ as 
\begin{align}
\Var(\hat{\beta}) &= \Var(\bar{Y}(1)-\bar{Y}(0)). \label{eq:vbeta}
\end{align}
Let $\mathcal{Z}$ be the covariates causing non-homogeneity, we rewrite \eqref{eq:vbeta} as
\begin{align}
\Var(\hat{\beta}) &= \Var_\mathcal{Z}\left(\E((\bar{Y}(1) - \bar{Y}(0))|\mathcal{Z})\right) + \E_\mathcal{Z}\left(\Var((\bar{Y}(1) - \bar{Y}(0))|\mathcal{Z})\right), \label{eq:21}\\
&=\Var_\mathcal{Z}\left(\E(\bar{Y}(1)|\mathcal{Z}) - \E(\bar{Y}(0)|\mathcal{Z})\right) + \E_\mathcal{Z}\left(\Var(\bar{Y}(1)|\mathcal{Z}) + \Var(\bar{Y}(0)|\mathcal{Z})\right). \label{eq:22}
\end{align}
\eqref{eq:22} uses linearity of expectation and randomization.

For fixed covariates the experimental units are identical. Therefore, $\E(\bar{Y}(x)|\mathcal{Z}) = Y(x)|\mathcal{Z}, x=0,1$, and $\Var(\bar{Y}(x)|\mathcal{Z}) = \frac{1}{n_{x,z}}\Var(Y(x)|\mathcal{Z}), x=0,1$. Therefore,
\begin{align}
\Var(\hat{\beta}) &= \Var_\mathcal{Z}\left(\E(Y(1)|\mathcal{Z}) - \E(Y(0)|\mathcal{Z})\right) + \E_\mathcal{Z}\left(\frac{\Var(Y(1)|\mathcal{Z})}{n_{1,z}} + \frac{\Var(Y(0)|\mathcal{Z})}{n_{0,z}}\right)
\end{align}
Define $\beta(\mathcal{Z}):= \E(Y(1)|\mathcal{Z}) - \E(Y(0)|\mathcal{Z})$. Therefore,
\begin{align}
\Var(\hat{\beta}) &= \Var_\mathcal{Z}\left(\beta(\mathcal{Z})\right) + \E_\mathcal{Z}\left(\frac{\Var(Y(1)|\mathcal{Z})}{n_{1,z}} + \frac{\Var(Y(0)|\mathcal{Z})}{n_{0,z}}\right)\\
&= \E_\mathcal{Z}\left(\beta(\mathcal{Z}) - \E_\mathcal{Z}(\beta(\mathcal{Z}))\right)^2 + \E_\mathcal{Z}\left(\frac{\Var(Y(1)|\mathcal{Z})}{n_{1,z}} + \frac{\Var(Y(0)|\mathcal{Z}}{n_{0,z}})\right) \label{eq:25}\\
&= \E_\mathcal{Z}\left(\beta(\mathcal{Z}) - \beta)\right)^2 + \E_\mathcal{Z}\left(\frac{\Var(Y(1)|\mathcal{Z})}{n_{1,z}} + \frac{\Var(Y(0)|\mathcal{Z})}{n_{0,z}}\right) \label{eq:26}
\end{align}
\eqref{eq:26} uses $\E_\mathcal{Z}(\beta(\mathcal{Z}))= \E_\mathcal{Z}(\E(Y(1)|\mathcal{Z})) - \E_\mathcal{Z}(\E(Y(0)|\mathcal{Z})) = \E(Y(1))-\E(Y(0)) = \beta$.

\subsection{Expectation and Variance of $\hat{\beta}_{\mathcal{Z}}$} \label{app:proofs_hat_betaz}

We calculate the expectation of $\hat{\beta}_{\mathcal{Z}}$ as
\begin{align}
\E(\hat{\beta}_\mathcal{Z}) & = \E\left(\sum_{\mathcal{Z}} \hat{\mathbb{P}}(\mathcal{Z}) \left(\bar{Y}(1) | \mathcal{Z} - \bar{Y}(0) | \mathcal{Z} \right)\right),\\
& = \sum_{\mathcal{Z}}  \E \left(\hat{\mathbb{P}}(\mathcal{Z}) (\bar{Y}(1) | \mathcal{Z} - \bar{Y}(0) | \mathcal{Z}) \right),\\
& = \sum_{\mathcal{Z}}  \E \left(\hat{\mathbb{P}}(\mathcal{Z})\right) \left(\E(\bar{Y}(1) | \mathcal{Z}) - \E(\bar{Y}(0) | \mathcal{Z})\right),\\
& = \sum_{\mathcal{Z}}  \mathbb{P}(\mathcal{Z}) \left(\E(\bar{Y}(1) | \mathcal{Z}) - \E(\bar{Y}(0) | \mathcal{Z})\right),\\
&= \E_\mathcal{Z}\left(\E(\bar{Y}(1) | \mathcal{Z}) - \E(\bar{Y}(0) | \mathcal{Z})\right) = \beta.
\end{align}
%Therefore, the unbiasedness of $\hat{\beta}_{\mathcal{Z}}$ for $\beta$.

We next calculate the variance of $\beta_{\mathcal{Z}}$ as
\begin{align}
\Var(\hat{\beta}_\mathcal{Z}) &= \Var_\mathcal{Z}\left(\E\left(\sum_\mathcal{Z}(\bar{Y(1)}|\mathcal{Z}-\bar{Y(0)}|\mathcal{Z})\hat{\P}(Z)\right)\right) + \E_\mathcal{Z}\left(\Var\left(\sum_\mathcal{Z}(\bar{Y(1)}|\mathcal{Z}-\bar{Y(0)}|\mathcal{Z})\hat{\P}(Z)\right)\right), \label{eq:32}\\
&= \Var_\mathcal{Z}\left(\sum_\mathcal{Z}\E(\bar{Y(1)}|\mathcal{Z}-\bar{Y(0)}|\mathcal{Z})\hat{\P}(Z)\right) + \E_\mathcal{Z}\left(\sum_\mathcal{Z}\Var(\bar{Y(1)}|\mathcal{Z}-\bar{Y(0)}|\mathcal{Z})\hat{\P}(Z)\right), \label{eq:33} \\
&= \Var_\mathcal{Z}\left(\sum_\mathcal{Z}\E(\bar{Y}(1) | \mathcal{Z}) - \E(\bar{Y}(0) | \mathcal{Z})\hat{\P}(Z)\right) + \E_\mathcal{Z}\left(\sum_\mathcal{Z}\left(\Var(\bar{Y}_{1}|\mathcal{Z})+\Var(\bar{Y}_{0}|\mathcal{Z})\right)\hat{\P}(Z)\right), \label{eq:34}
\end{align}
\eqref{eq:33} and \eqref{eq:34} use linearity of expectation and randomization.

%\newpage

For fixed covariates the experimental units are identical. Therefore, $\E(\bar{Y}(x)|\mathcal{Z}) = \E(Y(x)|\mathcal{Z}), x=0,1$, and $\Var(\bar{Y}(x)|\mathcal{Z}) = \frac{1}{n_{x,z}}\Var(Y(x)|\mathcal{Z}), x=0,1$. Therefore,
\begin{align}
\Var(\hat{\beta}_\mathcal{Z}) &= \Var_\mathcal{Z}\left(\sum_\mathcal{Z}\beta(Z)\hat{\mathbb{P}}(\mathcal{Z})\right) + \E_\mathcal{Z}\left(\sum_\mathcal{Z}\left(\frac{\Var(Y(1)|\mathcal{Z})}{n_{1,z}}+\frac{\Var(Y(0)|\mathcal{Z})}{n_{0,z}}\right)\hat{\mathbb{P}}(\mathcal{Z})\right),\\
&= \E_\mathcal{Z}\left(\sum_\mathcal{Z}\beta(Z)\hat{\mathbb{P}}(\mathcal{Z}) - \E_\mathcal{Z}\left(\sum_\mathcal{Z}\beta(Z)\hat{\mathbb{P}}(\mathcal{Z})\right) \right)^2 + \sum_\mathcal{Z}\frac{\Var(Y(1)|\mathcal{Z})}{n_{1,z}}\E_\mathcal{Z}(\hat{\mathbb{P}}(\mathcal{Z})) \nonumber\\
&\hspace{.3in}+ \sum_\mathcal{Z}\frac{\Var(Y(0)|\mathcal{Z})}{n_{0,z}}\E_\mathcal{Z}(\hat{\mathbb{P}}(\mathcal{Z})),
\end{align}

\begin{align}
&= \E_\mathcal{Z}\left(\sum_\mathcal{Z}\beta(Z)\hat{\mathbb{P}}(\mathcal{Z}) - \beta \right)^2 + \sum_\mathcal{Z}\frac{\Var(Y(1)|\mathcal{Z})}{n_{1,z}}\mathbb{P}(Z) + \sum_\mathcal{Z}\frac{\Var(Y(0)|\mathcal{Z})}{n_{0,z}}\mathbb{P}(Z),\\
&= \E_\mathcal{Z}\left(\sum_\mathcal{Z}\beta(Z)\hat{\mathbb{P}}(\mathcal{Z}) - \beta \right)^2 + \E_\mathcal{Z}\left(\frac{\Var(Y(1)|\mathcal{Z})}{n_{1,z}}\right) + \E_\mathcal{Z}\left(\frac{\Var(Y(1)|\mathcal{Z})}{n_{1,z}}\right),\\
&= \E_\mathcal{Z}\left(\frac{\Var(Y(1)|\mathcal{Z})}{n_{1,z}} +\frac{\Var(Y(0)|\mathcal{Z})}{n_{0,z}} \right) + \E_\mathcal{Z}\left(\sum_\mathcal{Z}\beta(Z)\hat{\mathbb{P}}(\mathcal{Z}) - \beta \right)^2.
\end{align}

%\newpage

\section{Details of the Experimental Evaluations} \label{app:exp_details}

In this section, we provide details of the experimental evaluations.

\subsection{Variable Definitions}

For the semi-Markovian causal graph, $\mathcal{G}$ given in \cref{fig:final_implementation1G}, we define the treatment and response as follows.
\begin{align*}
\text{Drug} &= 
\begin{cases} 
    1, & \text{if an individual receives the drug}, \\
    0, & \text{otherwise,}\\
\end{cases}\\
\text{Blood-Pressure} &= 
\begin{cases} 
    1, & \text{if an individual has high blood pressure}, \\
    0, & \text{otherwise,}
\end{cases}
\end{align*}

We define the observed covariates as follows.
\begin{align*}
\text{Food Habits} &= 
\begin{cases} 
    1, & \text{if an individual's food habits are unhealthy}, \\
    0, & \text{otherwise,}\\
\end{cases}\\
\text{Blood-Sugar} &= 
\begin{cases} 
    1, & \text{if an individual has high blood sugar}, \\
    0, & \text{otherwise,}\\
\end{cases}\\
\text{Age} &= 
\begin{cases} 
    1, & \text{if an individual is above 40 years of age}, \\
    0, & \text{otherwise,}
\end{cases}\\
\text{Alcohol} &= 
\begin{cases} 
    1, & \text{if an individual consumes alcohol}, \\
    0, & \text{otherwise,}
\end{cases}\\
\text{Cholesterol} &= 
\begin{cases} 
    1, & \text{if an individual has high cholesterol}, \\
    0, & \text{otherwise,}
\end{cases}\\
\end{align*}

\begin{align*}
\text{Anxiety} &= 
\begin{cases} 
    1, & \text{if an individual has anxiety}, \\
    0, & \text{otherwise,}
\end{cases}\\
\text{Sleep Quality} &=
\begin{cases} 
    1, & \text{if an individual's sleep quality is bad}, \\
    0, & \text{otherwise,}
\end{cases}\\
\text{Palpitations} &= 
\begin{cases} 
    1, & \text{if an individual suffers from palpitations}, \\
    0, & \text{otherwise,}
\end{cases}\\
\text{Strenuous Activity} &= 
\begin{cases} 
    1, & \text{if an individual does strenuous activity}, \\
    0, & \text{otherwise.}
\end{cases}
\end{align*}
The graph $\mathcal{G}$ in Figure \ref{fig:final_implementation1G} has three unobserved variables. We label them as $\text{Anxiety} \dashleftarrow U_1 \dashrightarrow \text{Alcohol}$, $\text{Anxiety} \dashleftarrow U_2 \dashrightarrow \text{Blood-Pressure}$, and $\text{Blood-Pressure} \dashleftarrow U_3 \dashrightarrow \text{Cholesterol}$. We consider each of $U_1, U_2$, and $U_3$ to be identically and independently distributed as $\text{Uniform}(-0.15,0.15)$.

\subsection{Probability Distributions}

For the observed variables, we consider the following probability distributions. For $x \in \{0,1\}$,
\begin{align*}
    \P(\text{Food Habits} = x) &= 0.5,\\
    \P(\text{Age} = x) &= 0.5, \\
    \P(\text{Alcohol} = x) &= \text{Food-Habits}[(0.3+U_1)(1-x) + (0.7-U_1)x] \\ 
    &\qquad{} + (1-\text{Food-Habits})[(0.3+U_1)(1-x) + (0.7-U_1)x], \\
    \P(\text{Cholesterol} = x) &= \text{Age}[(0.3+U_3)(1-x) + (0.7-U_3)x] \\ 
    &\qquad{} + (1-\text{Age})((0.3+U_3)(1-x) + (0.7-U_3)x),\\
    \P(\text{Anxiety} = x) &= \text{Drug}[(0.3+U_2)(1-x) + (0.7-U_2)x] \\ 
    &\qquad{} + (1-\text{Drug})((0.3+U_2)(1-x) + (0.7-U_2)x),\\
    \P(\text{Sleep Quality} = x) &= \text{Anxiety}[(0.3+U_2)(1-x) + (0.7-U_2)x] \\ 
    &\qquad{} + (1-\text{Anxiety})((0.3+U_2)(1-x) + (0.7-U_2)x),\\
    \P(\text{Blood-Pressure} = x) &= W[(0.3+U_1-U_2)(1-x) + (0.7-U_1+U_2)x] \\ 
    &\qquad{} + (1-W)[(0.3-U_1+U_2)(1-x) + (0.7+U_1-U_2)x],
\end{align*}
where $W = \text{Alcohol} \times \text{Cholesterol} \times \text{Anxiety} \times \text{Sleep-Quality}$.

\subsection{Baselines}
%To the best of our knowledge, ours is the first general algorithm to obtain a stable causal blocking set. In the past \citep{cinelli2020crash}, there have been discussions about obtaining the optimal blocking sets for specific graphs but a general algorithm did not exist. Hence, for empirical comparisons, we construct the following baselines by choosing some representative subsets of the observed covariates. %We explain our reasoning behind these choices as follows. 

For empirical comparisons, we constructed the following baselines by choosing some representative subsets of the observed covariates.
\begin{enumerate}
    \item $\phi$: No blocking, i.e., a completely randomized design. We use this as a baseline to compare if our blocking set performs at least better than no blocking.
    \item \{Food Habits\}: Food Habits affect the response (Blood-Pressure) through a mediator, Alcohol. We use \{Food Habits\} as the blocking set to see how using only a non-parent ancestor of the response compares with our blocking set.
    \item \{Food Habits, Alcohol\}: As discussed above Food Habits affect the response through Alcohol. We use \{Food Habits, Alcohol\} as the blocking set to see how only this parent and non-parent ancestor pair compare with our blocking set.
    \item \{Food Habits, Alcohol, Cholesterol\}: This subset includes the above parent and non-parent ancestor pair, \{Food Habits, Alcohol\} and one more parent of the response, Cholesterol while leaving out Cholesterol's parent, Age. We use \{Food Habits, Alcohol, Cholesterol\} as the blocking set to see how leaving out Age in the presence of a latent variable path compares with our blocking set.
\end{enumerate}

\subsection{Simulation and Estimation}
We performed the experiment with no blocking by randomizing the treatment (Drug) 100 times. For performing the experiment with blocking, we first created blocks using the chosen set of covariates and then randomized the treatment within each block. Using the above probability distributions, we generated/simulated 100 samples of the response (Blood-Pressure). We calculated the causal effect of the drug on blood pressure as the average of the response. We repeated the above experiment 10 times. The estimated causal effect and the variability in the response (averaged across 10 runs) for different choices of blocking sets are provided in Table \ref{tab:causal_effect}. 

\subsection{Testing for Statistical Significance}
We tested for the statistical significance of all pairwise differences between the estimates of the causal effect using different blocking sets using a $t$-test. The $p$-values are provided in \cref{tab:causal_effect_p_values}. From \cref{tab:causal_effect_p_values}, we observe that the $p$-values for all pairwise differences in causal effect estimates are quite high. Hence, we conclude that these differences are not significant. This confirms the idea that blocking does not alter the causal effect estimates. We tested for the statistical significance of all pairwise ratios of the estimates of the variability in the response using different blocking sets using an $F$-test. The $p$-values are provided in \cref{tab:var_response_p_values}. From the last column of \cref{tab:var_response_p_values}, we observe that the $p$-values for comparing the variance using our blocking set with other blocking sets are fairly low (maximum being 0.2567). We can say that the differences in variance are significant at a 25.67\% level of significance. %We understand that the level of significance is typically taken around 10-15\%. Hence, to be sure, we repeated our experiment 10 times. The above $p$-values are based on the average effect and variance across 10 different runs.

%Finally, we estimated the causal effect and variability in the response for those different choices of blocking sets.

\begin{table*}[t] 
    \centering 
    \caption{$p$-values for the difference between causal effect estimates.}  \label{tab:causal_effect_p_values}
	\begin{tabular}{l p{2cm} p{2cm} p{2cm} p{2.5cm}}
		\hline \hline
		  $p$-value & \{Food Habits\} & \{Food Habits, Alcohol\} & \{Food Habits, Alcohol, Cholesterol\} & \{Food Habits, Alcohol, Cholesterol, Age\} \\
            \hline
            $\phi$ & 0.9511 & 0.9389 & 0.8196 & 0.7103 \\
            \{Food Habits\} & & 0.9877 & 0.7723 & 0.7565 \\
            \{Food Habits, Alcohol\} & & & 0.7606 & 0.7682 \\
            \{Food Habits, Alcohol, Cholesterol\} & & & & 0.5490 \\
	    \hline
	    \hline
	\end{tabular}
\end{table*}

\begin{table*}[t] 
    \centering 
    \caption{$p$-values for the ratio of variance estimates.}  \label{tab:var_response_p_values}
	\begin{tabular}{l p{2cm} p{2cm} p{2cm} p{2.5cm}}
		\hline \hline
		  $p$-value & \{Food Habits\} & \{Food Habits, Alcohol\} & \{Food Habits, Alcohol, Cholesterol\} & \{Food Habits, Alcohol, Cholesterol, Age\} \\
            \hline
            $\phi$ & 0.4738 & 0.4074 & 0.3818 & 0.1701 \\
            \{Food Habits\} & & 0.4331 & 0.4071 & 0.1872 \\
            \{Food Habits, Alcohol\} & & & 0.4735 & 0.2385 \\
            \{Food Habits, Alcohol, Cholesterol\} & & & & 0.2567 \\
	    \hline
	    \hline
	\end{tabular}
\end{table*}

\section{Future Direction to use the Proposed Algorithm for Different Real-World Applications} \label{app:future}

In this section, we discuss some future directions to use the proposed algorithm for different real-world applications. The proposed algorithm efficiently obtains a stable and variance-minimizing set of covariates to be used for forming blocks while performing random experiments. Random experiments are commonly used to study the causal effect. For instance in answering the following questions. Does a new seed variety help in improving crop yield? Does a new medicine help in alleviating muscle pain? Does offering discounts help in product adoption? Does recommending a set of products improve the user-satisfaction? From a reinforcement learning standpoint, an intervention like a new seed variety, a new medicine, a new promotional offer, a new product recommendation, etc. can be seen as an action, and the response like the crop yield, reduction in muscle pain, number of product adoptions, user-satisfaction, etc. can be regarded as the reward. A common question in reinforcement learning is to find a reward-maximizing action that is equivalent to finding good interventions in randomized experiments. Hence, the algorithm proposed in this paper can be used in the efficient and low-variance estimation of causal effects to identify good actions/interventions in different real-world applications. We are interested in the applications of the proposed algorithm to offline \citep{kempe2003maximizing, umrawal2022community} and online \citep{agarwal2022stochastic,nie2022explore} social influence maximization -- the task of selecting a set of individuals on a social network to be given some discounts to cause a large cascade of further product adoptions.
\end{document}